\documentclass[11pt]{article}

\usepackage{tikz}
\usetikzlibrary{fit,positioning,calc}
\usepackage{amsmath}  
\usepackage{amssymb}
\usepackage{bbold}
\usepackage{geometry}
\usepackage{graphicx,psfrag,epsf}
\usepackage{enumerate}
\usepackage[colorlinks=false]{hyperref}
\usepackage{cleveref}
\usepackage{natbib}
\usepackage{booktabs}
\usepackage{array}
\usepackage{float}

\begin{document}

\def\spacingset#1{\renewcommand{\baselinestretch}%
{#1}\small\normalsize} \spacingset{1}

\title{\bf Bayesian Network-Response Regression}

\author{
Lu Wang$^1$,  Daniele Durante$^2$, Rex E.~Jung$^3$, and David B.~Dunson$^1$ \bigskip \\
$^1$Department of Statistical Science, Duke University, Durham, NC 27708, USA \\
$^2$Department of Statistical Sciences, University of Padova, Padova 241, 35121, Italy \\
$^3$Department of Psychology, University of New Mexico, Albuquerque, NM 87131, USA \\
\bigskip \small{Email: \texttt{rl.wang@duke.edu}, \texttt{durante@stat.unipd.it}, \texttt{rex.jung@runbox.com}, \texttt{dunson@duke.edu}}
}
\date{}
\maketitle

\spacingset{1.45}
\begin{abstract}
There is increasing interest in learning how human brain networks vary as a function of a continuous trait, but flexible and efficient procedures to accomplish this goal are limited. We develop a Bayesian semiparametric model, which combines low-rank factorizations and flexible Gaussian process priors to learn changes in the conditional expectation of a network-valued random variable across the values of a continuous predictor, while including subject-specific random effects. The formulation leads to a general framework for inference on changes in brain network structures across human traits, facilitating borrowing of information and coherently characterizing uncertainty. We provide an efficient Gibbs sampler for posterior computation along with simple procedures for inference, prediction and goodness-of-fit assessments. The model is applied to learn how human brain networks vary across individuals with different intelligence scores.  Results provide interesting insights on the association between intelligence and brain connectivity, while demonstrating good predictive performance.
\end{abstract}


\section{Introduction}
\label{intro}
We are motivated by recent advances in neuroimaging of structural interconnections among anatomical regions in the human brain.  Our focus is on learning how brain structural connectivity networks  -- also known as connectomes -- vary across individuals, and the extent to which such variability is associated with differences in human cognitive traits.

\begin{figure}[t]
\centering
\includegraphics[scale=0.65]{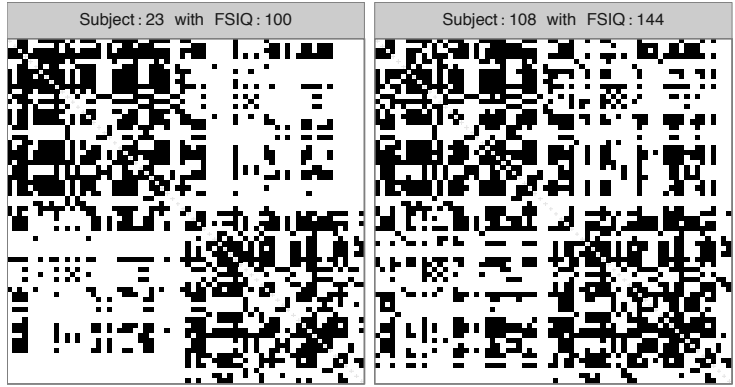} 
\caption{Binary adjacency matrices for two selected subjects. Black cells denote presence of a connection. White cells denote no connection. FSIQ is an IQ score.}
\label{fig:1}
\end{figure}

In our application, brain networks are estimated exploiting structural magnetic resonance imaging and diffusion tensor imaging to obtain a $V\times V$ symmetric adjacency matrix $A_{i}$ for each subject $i=1,\dots,n$. Each cell $[vu]$ in the matrix corresponds to a pair of brain regions, with $A_{i[vu]}=A_{i[uv]}=1$ if there are fibers connecting brain regions $v$ and $u$ in subject $i$, and $A_{i[vu]}=A_{i[uv]}=0$ otherwise. There are $V=68$ regions in our study \citep{Desikan:2006aa} equally divided in the left and right hemisphere. Refer to Figure~\ref{fig:1} for an example of the available data.

There has been an increasing focus on using brain imaging technologies to better understand the neural pathways underlying human traits, ranging from personality to cognitive abilities and mental disorders  \citep{stam2014modern}. Our aim is to develop flexible procedures to improve understanding of how the brain structural connectivity architecture varies in relation to a trait of interest $x_{i} \in \Re$ measured for each subject $i=1, \ldots, n$.  In our  application this trait represents a measure of intelligence available via the FSIQ (Full Scale Intelligence Quotient) score \citep{jung2007parieto}. 

Network data are challenging to analyze because they require not only dimensionality reduction procedures to effectively deal with the large number of pairwise relationships,  but also flexible formulations to account for the topological structures of the network. Current literature addresses these goals only for a single network observation. Notable examples include exponential random graph models \citep[e.g.][]{fra_1986}, and factorizations covering stochastic block models \citep{now_2001}, mixed membership stochastic block models \citep{air_2008} and latent space models \citep{hof_2002}. These procedures reduce dimensionality, incorporate network properties and have been generalized to accommodate regression settings in which there are response variables and network-specific predictors associated with every node $v=1, \ldots, V$ in a single network \citep[e.g.][]{malley_2008, hoch_2007}.  This type of network regression is fundamentally different from our interest in relating a network $A_i$ specific to individual $i$ to a corresponding predictor $x_i$, for $i=1,\ldots,n$.

In relating the network $A_i$ to a specific trait $x_i$, a common strategy in neuroscience is to estimate a separate logistic regression for each pair of brain regions to learn changes in their connectivity with the predictor. However, as discussed in \citet{Simp_2013}, such massive univariate edge-based studies do not incorporate dependence in connectivity patterns, and therefore  ignore relevant wiring mechanisms in the brain architecture. This has motivated an increasing interest in how topological characteristics of a complex network change as a function of a trait \citep[e.g.][]{rubinov2010complex}. Typical procedures address this aim by computing a set of summary measures for each network --  e.g. network density, transitivity, average path length, assortativity -- and enter these statistics as responses in a regression model \citep[e.g.][]{van_2009,wu_2013}. However, reducing the rich network data to a subset of summary measures can discard important information about the brain connectivity and underestimate its changes across the trait. 

With these issues in mind, we develop a network--response regression model, which considers the brain network $A_i$ as an object-type response variable having conditional expectation changing flexibly with $x_i$.  To our knowledge, there is no literature addressing this problem, though there are a rich variety of methods for characterizing dynamic changes in a time-specific network $A_t$, with \citet{holl_1977}, \citet{xing_2010}, \citet{sewell_2015} considering discrete equally-spaced times and  \citet{durante_2014} providing a continuous-time formulation.  Although the later article provides a useful building block, the time-series setting is fundamentally different from the regression case we consider, motivating careful modifications to incorporate subject-specific variability and other relevant structure.

There is an increasing availability of data motivating network--response regression models.  We propose a Bayesian semiparametric formulation that reduces dimensionality and efficiently exploits network information via a flexible latent space representation, with the latent coordinates of the brain regions varying both systematically -- according to a trait of the individual, such as FSIQ -- and randomly -- due to unobserved traits or measurement errors -- across individuals. This formulation allows coherent inference at different scales, including global changes in network topological structures and local variations in edge probabilities.

The paper is organized as follows. In Section \ref{method} we focus on model formulation, prior specification and posterior computation. A simulation study is examined in Section \ref{simu}. In Section \ref{app} our model is applied to learn changes in the brain network with the FSIQ score, showing improved performance in inference, edge prediction and uncertainty quantification.

\section{Methods}\label{method}
Let ${A}_{i}$ denote the binary adjacency matrix characterizing the undirected brain network with no self-relationships for the subject $i$, and $x_i$ the corresponding trait value, for every $i=1, \ldots, n$. As self-relationships are not of interest and ${A}_{i}$ is symmetric, we model $A_{1}, \ldots, A_{n}$ by defining a probabilistic generative mechanism for the vectors  $\mathcal{L}({A}_{1}), \ldots, \mathcal{L}({A}_{n})$, with $\mathcal{L}({A}_{i})=(A_{i[21]}, A_{i[31]}, \ldots , A_{i[V1]},$ $ A_{i[32]}, \ldots A_{i[V(V-1)]})^\intercal$ the vector encoding the lower triangular elements of $A_{i}$, which uniquely characterize $A_i$. Hence, $\mathcal{L}({A}_{i})$ is a vector of binary elements $\mathcal{L}({A}_{i})_l \in \{0,1\}$, $l=1, \ldots, V(V-1)/2$,  encoding the absence or presence of an edge among the $l$th pair of brain regions in subject $i$.

Based on our notation, developing a regression model for a network-valued response translates into statistical modeling of how a vector of binary data changes across the values of a trait of interest. However, it is important to explicitly incorporate the network structure of our data. In fact, networks are potentially characterized by specific underlying topological patterns which induce  dependence among the edges within each brain. As a result, by carefully accommodating the network structure in modeling of $\mathcal{L}({A}_{1}), \ldots, \mathcal{L}({A}_{n})$, one might efficiently borrow information within each  $\mathcal{L}({A}_{i})$ and across the trait $x_i$, while reducing dimensionality and inferring specific network properties along with their changes across $x_i$.

\subsection{Model formulation}
In modeling of  $\mathcal{L}({A}_{1}), \ldots, \mathcal{L}({A}_{n})$ we look for a representation which can flexibly characterize variability across individuals in brain connectivity, while accommodating network structure within each brain and learning changes with the trait $x_i$. Individual variability \citep[e.g][]{Mueller:2013aa} and specific network structures \citep[e.g][]{bullmore2009complex} have been shown to substantially affect the functioning of networked brain systems, with these systems often varying with human traits \citep[e.g][]{stam2014modern}. 

Consistent with the above goals and letting $\mathcal{L}(\mathcal{A}_i)$ denote the random variable associated with the brain network of subject $i$, we characterize individual variability by assuming the edges among pairs of brain regions  are conditionally independent Bernoulli variables, given a subject-specific edge probability vector $\pi^{(i)}=\{\pi_1^{(i)}, \ldots, \pi_{V(V-1)/2}^{(i)} \}^\intercal$,
\begin{eqnarray}
\mathcal{L}(\mathcal{A}_i)_l \mid \pi_l^{(i)} \sim \mbox{Bern}\{\pi_l^{(i)}\}, \quad \pi_l^{(i)}=\mbox{pr}\{\mathcal{L}(\mathcal{A}_i)_l=1\},
\label{eq1}
\end{eqnarray} 
independently for each pair $l=1, \ldots, V(V-1)/2$ and $i=1, \ldots, n$. Equation~\eqref{eq1} incorporates individual variability, but fails to account for two key sources of information in our data. In fact, we expect dependence between the edge probabilities in each $\pi^{(i)}$ due to the network topology. Moreover, it is reasonable to expect that subjects with similar traits will have comparable brain networks. To incorporate these structures, we define the edge probabilities as a function of subject-specific node positions in a latent space, with these positions centered on a higher-level mean which changes with $x_i$. Specifically, letting $l$ denote the pair of brain regions $v$ and $u$, $v>u$, we first borrow information within each $\pi^{(i)}$ by defining
\begin{eqnarray}
\mbox{logit}\{\pi_l^{(i)}\}= Z_l+\sum_{r=1}^R Y^{(i)}_{vr}Y^{(i)}_{ur},
\label{eq2}
\end{eqnarray} 
for each $l=1, \ldots, V(V-1)/2$ and $i=1, \ldots, n$. In \eqref{eq2}, $Z_l\in \Re$ is a similarity measure for the $l$th pair of regions shared among all individuals, whereas $Y^{(i)}_{vr} \in \Re$ and $Y^{(i)}_{ur} \in \Re$ denote the $r$th coordinate of the brain regions $v$ and $u$ for subject $i$, respectively. This construction has also an intuitive interpretation. In fact,  $Y_{vr}^{(i)}$ may measure the propensity of brain region $v$ towards the $r$th  cognitive function in subject $i$. According to  \eqref{eq2}, if regions $v$ and $u$ have propensities in the same direction, they will have a high chance $\pi_l^{(i)}$ to be connected. Moreover, embedding the brain regions in a lower-dimensional space via \eqref{eq2} allows dimensionality reduction, and can accommodate several topological properties \citep{hoff2009hierarchical}.

\subsection{Prior specification}
To conclude our Bayesian specification, we choose priors for the shared parameters $Z_l$, $l=1, \ldots, V(V-1)/2$, and the subject-specific latent coordinates $Y^{(i)}_{vr}$ for each $v=1, \ldots, V$, $r=1, \ldots, R$ and $i=1, \ldots, n$. These priors are defined to facilitate simple posterior computation, favor borrowing of information between different individuals and allow the latent coordinates to smoothly change across the values of the predictor. Subjects with similar traits are expected to have comparable brain networks. We incorporate this structure by centering the prior for the subject-specific latent coordinates on a higher-level mean smoothly changing with the trait of interest. Then, by updating this prior with the likelihood provided by the data we expect the posterior to flexibly account for possible deviations from prior assumptions, including allowance  for uninformative traits. 

Consistent with the above considerations, we let
\begin{eqnarray}
Z_{l} \sim \mbox{N}(\mu_z,\sigma_z^2),
\label{eq3}
\end{eqnarray} 
independently for $l=1, \ldots, V(V-1)/2$, and
\begin{eqnarray}
Y^{(i)}_{vr} \sim \mbox{N}\{\mu_{vr}(x_i),1\},
\label{eq4}
\end{eqnarray} 
independently for every $v=1, \ldots, V$, $r=1, \ldots, R$ and $i=1, \ldots, n$. We set the prior variance of $Y^{(i)}_{vr}$ at 1 because we observed indistinguishable results when replacing \eqref{eq4} with a Student-$t$ distribution. Hence we maintain the Gaussian prior \eqref{eq4} to keep the model parsimonious and computationally tractable. To accommodate systematic deviations, we incorporate mean functions $\mu_{vr}(\cdot)$ characterizing changes in the $r$th latent coordinate of the $v$th brain region with the trait of interest. 

In modeling $\mu_{vr}(\cdot)$, we could consider a Gaussian process (GP) prior for each $v=1, \ldots, V$ and $r=1, \ldots, R$. However, as the number of nodes increases, we face scalability issues.  To reduce dimensionality, we define each $\mu_{vr}(\cdot)$ as a linear combination of a smaller number of  dictionary functions $W_{kr}(\cdot)$, $k=1, \ldots, K<V$ and $r=1, \ldots, R$,
\begin{eqnarray}
\mu_{vr}(\cdot) & = & \sum_{k=1}^{K}G_{vk}W_{kr}(\cdot),
\label{eq5}
\end{eqnarray}
for each $v=1, \ldots, V$ and $r=1, \ldots, R$, where $G_{vk}$, for $v=1, \ldots, V$ and $k=1, \ldots, K$, are coefficients common to all the subjects. Factorization \eqref{eq5} further reduces the number of unknown functions from $V\times R$ to $K\times R$, where $K$ is typically smaller than $V$. Factorizations \eqref{eq2} and \eqref{eq5} are not unique; however, we avoid identifiability constraints as the focus of our inference is not on latent coordinates but on how the overall network structure varies systematically with traits and randomly across individuals.  Such inferences can be accomplished via appropriate functionals of the edge probabilities in $\pi^{(i)}$, as we will illustrate.

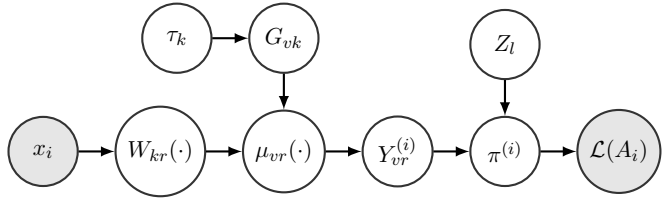
\begin{figure}[t]
\centering
\begin{tikzpicture}[scale=0.8, transform shape]
\tikzstyle{main}=[circle, minimum size = 11.5mm, thick, draw =black!80, node distance = 6mm]
\tikzstyle{connect}=[-latex, thick]
\tikzstyle{box}=[rectangle, draw=black!100]
  \node[main, fill = black!10] (A) {$\mathcal{L}({A}_i)$ };
    \node[main] (pi) [left=of A] {$\pi^{(i)}$};
        \node[main] (z) [above=of pi] {$Z_l$};
                \node[main] (y) [left=of pi] {$Y_{vr}^{(i)}$};
                                \node[main] (mu) [left=of y] {$\mu_{vr}(\cdot)$};
\node[main] (g) [above=of mu] {$G_{vk}$};
\node[main] (tau) [left=of g] {$\tau_k$};
\node[main] (w) [left=of mu] {$W_{kr}(\cdot)$};
\node[main, fill = black!10] (x) [left=of w] {$x_i$};
\path        (pi) edge [connect] (A);
 \path        (z) edge [connect] (pi);
 \path        (y) edge [connect] (pi);
  \path        (mu) edge [connect] (y);
 \path        (g) edge [connect] (mu);
 \path        (tau) edge [connect] (g);
 \path        (w) edge [connect] (mu);
\path        (x) edge [connect] (w);
\end{tikzpicture}
\caption{Graphical representation of our Bayesian network--response regression model. }
\label{fig:graphic}
\end{figure}

In choosing priors for the components in factorization  \eqref{eq5}, we first let
\begin{eqnarray}
W_{kr}(\cdot)\sim\mbox{GP}(0,c)
\label{eq6}
\end{eqnarray}
independently for each $k=1, \ldots, K$ and $r=1, \ldots R$, with $c$ the squared exponential correlation function $c(x_i,x_j)=\exp\{-\kappa( x_i-x_j)^{2}\}$, $\kappa>0$. To allow adaptive deletion of unnecessary dictionary functions, we incorporate a shrinkage effect in the prior for the coefficients $G_{vk}$, $v=1, \ldots, V$ and $k=1, \ldots, K$, letting
\begin{eqnarray}
G_{vk}&\sim& \mbox{N}(0,\tau_{k}^{-1}), \quad v=1,\dots,V, \ k=1,\dots,K, \label{eq7}\\
\tau_{k}&\sim& \mbox{Ga}\{aq^{3(k-1)},q^{2(k-1)}\}, \ q>1.
\label{eq8}
\end{eqnarray}
In \eqref{eq7}, $\tau_{k}$ provides a global column-wise shrinkage effect on $G_{vk}$. High values of $\tau_{k}$ force the prior for $G_{vk}$ to be concentrated around zero {\em a priori}, deleting the effect of the corresponding dictionary function $W_{kr}(\cdot)$ in factorization \eqref{eq5}. Equation \eqref{eq8} is carefully defined to allow this shrinkage effect to be increasingly strong as $k$ grows. A graphical representation of our hierarchical model formulation is provided in Figure~\ref{fig:graphic}. 

\subsection{Posterior computation}
Given priors defined in equations \eqref{eq3}--\eqref{eq8}, posterior computation for model \eqref{eq1} with subject-specific edge probabilities factorized as in \eqref{eq2}, proceeds via a simple Gibbs sampler leveraging  P\'olya-Gamma data augmentation \citep{pol_2013}, which allows conjugate inference in Bayesian logistic regression. We summarize below the main steps of the MCMC routine. Step-by-step derivations  are provided in the Supplementary Material.
\begin{itemize}
\item{Update each augmented data $\omega^{(i)}_l$, $l=1, \ldots, V(V-1)/2$, $i=1, \ldots, n$, from its full conditional P\'olya-Gamma distribution.}
\item{Given the data $\{\mathcal{L}({A}_{i}), x_i\}$,  $i=1, \ldots,n$, the latent coordinates' matrix $Y^{(i)}$, $i=1, \ldots,n$, and the P\'olya-Gamma augmented data $\omega^{(i)}_l$,  $l=1, \ldots, V(V-1)/2$, $i=1, \ldots, n$, the full conditionals for $Z_l$, $l=1, \ldots, V(V-1)/2$ are Gaussian distributions.}
\item{In updating the subject-specific coordinates' matrix $Y^{(i)}$, for each $i=1, \ldots, n$, we block sample the rows of $Y^{(i)}$ in turn conditionally on the rest of the matrix and the parameters $G_{vk}$, $v=1, \ldots, V$, $k=1, \ldots, K$, $W_{kr}(x_i)$, $k=1, \ldots, K$, $r=1,\ldots, R$. This approach allows rewriting the model \eqref{eq1}--\eqref{eq2} as a Bayesian logistic regression for which the P\'olya-Gamma data augmentation scheme guarantees conjugate Gaussian full conditionals.}
\item{Given the coordinates' matrix $Y^{(i)}$, $i=1, \ldots, n$ and the traits $x_i$, $i=1, \ldots, n$, updating for the parameters  $G_{vk}$, $v=1, \ldots, V$, $k=1, \ldots, K$ and the trajectories $W_{kr}(\cdot)$, $k=1, \ldots, K$, $r=1,\ldots, R$ at the observed trait values, proceed by exploiting the properties of GP priors and standard steps in Bayesian linear regression. Since the data are observed for a finite number of subjects, this step uses the multivariate Gaussian representation of the GP. However it is worth noticing that our  model is inherently semiparametric as the GP in \eqref{eq6} induces a prior on the infinite-dimensional space of  smooth functions. }
\item{Conditioned on  $G_{vk}$, $v=1, \ldots, V$, $k=1, \ldots, K$, the shrinkage parameters $\tau_k$, $k=1, \ldots, K$ are updated from their full conditional Gamma distributions.}
\item{To obtain each $\pi^{(i)}_l$,  $l=1, \ldots, V(V-1)/2$, $i=1, \ldots, n$ simply apply equation  \eqref{eq2} to the posterior samples of $Y^{(i)}$, for each $i=1, \ldots, n$ and $Z_l$, $l=1,\ldots, V(V-1)/2$.}
\item{Impute missing edges  $\mathcal{L}(A_i)_l$ from $\mathcal{L}(A_i)_l \mid \pi^{(i)}_l \sim  \mbox{Bern}\{\pi_l^{(i)}\}.$}
\end{itemize}

Obtaining  posterior samples for the subject-specific edge probabilities associated to missing edges is a key for prediction. Under our Bayesian procedure and recalling equation \eqref{eq1}, prediction of unobserved edges can be obtained by exploiting the mean of the posterior predictive distribution
\begin{align}
& \mbox{E}\{\mathcal{L}(\mathcal{A}_i)_l \mid\mathcal{L}({A}_1), \ldots, \mathcal{L}({A}_n) \} \nonumber \\
& = \mbox{E} \left[ \mbox{E}\{\mathcal{L}(\mathcal{A}_i)_l \mid \pi^{(i)}_l \} \mid \mathcal{L}({A}_1), \ldots, \mathcal{L}({A}_n) \right] \label{eq10} \\
& = \mbox{E} \{ \pi_l^{(i)} \mid \mathcal{L}({A}_1), \ldots, \mathcal{L}({A}_n) \nonumber \}, \quad l=1, \ldots, V(V-1)/2, 
\end{align}
for each possible missing edge in subject $i=1, \ldots, n$, where the last expectation coincides with the posterior mean of $\pi_l^{(i)}$. Note that we use standard font $\mathcal{L}({A})$ to define the observed vectors of edges and italics notation $\mathcal{L}(\mathcal{A})$ to denote the associated random variable.

\section{Simulation study}\label{simu}
We evaluate the performance of our methods on synthetic data simulated from a generating process different than our statistical model. Our goal is to assess whether the proposed methods are sufficiently flexible to learn global and local changes in brain connectivity structures, even when such variations arise from different generative mechanisms.

To accomplish the above goal, we simulate multiple brain networks $A_i$ with $V=20$ nodes and having predictors $x_i$ observed on a discrete grid $x_i\in \mathbb{X}= \{1, \ldots, 15\}$. In particular, for each unique predictor value, four networks $A_i$ are generated, for a total of $n=60$ subjects. To imitate the hemispheres and lobes in the brain, we define four node blocks $\mathbb{V}_{L_1}=\{1,\dots,5\}$, $\mathbb{V}_{L_2}=\{6,\dots,10\}$, $\mathbb{V}_{R_1}=\{11,\dots,15\}$ and $\mathbb{V}_{R_2}=\{16,\dots,20\}$. Nodes in $\mathbb{V}_{L_1}$ and $\mathbb{V}_{R_1}$ belong to the first lobe in the left and right hemisphere, respectively, whereas nodes in $\mathbb{V}_{L_2}$ and $\mathbb{V}_{R_2}$ belong to the second lobe in the left and right hemisphere, respectively.

\begin{figure*}[t]
\centering
\includegraphics[scale=0.58]{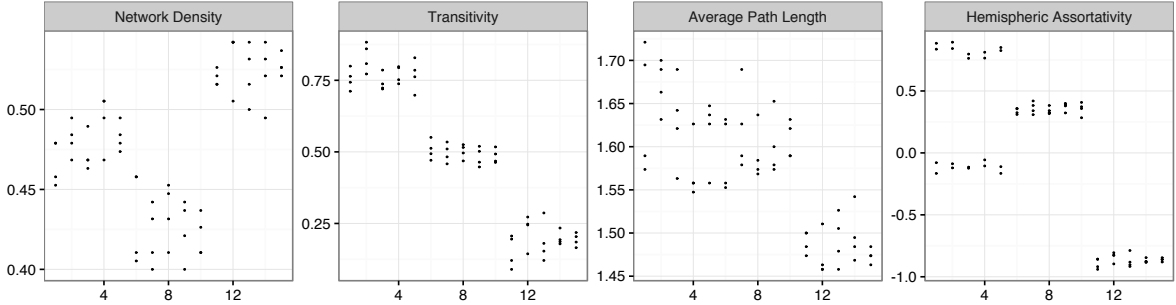}
\caption{Scatterplots of the network summary measures computed for the simulated networks $A_i$, $i=1, \ldots, n$ versus their corresponding predictor value $x_i$, $i=1, \ldots, n$.}
\label{fig:2}
\end{figure*}

In simulating the data $A_i$, we aim to incorporate different topological properties typically observed in human brain networks --- covering block-structures by hemispheres    and lobes, along with small-world architectures \citep[e.g][]{bullmore2009complex}. In particular, for each unique predictor value in $\mathbb{X}_1=\{1, \ldots, 5\}$, half of the subjects have high assortativity\footnote{In this context, the assortativity measures if nodes in the same block are more likely to be connected than nodes in different blocks.} by hemisphere, whereas the others have brains with high lobe assortativity. Subjects with an intermediate predictor value $x_i \in \mathbb{X}_2=\{6, \ldots, 10\}$ are characterized by brain networks having small-world behavior according to the \citet{watts_1998} model. Finally -- consistent with initial descriptive analyses of our data -- we increase the interhemispheric density and reduce the intrahemispheric connectivity in the brain networks of the subjects with high predictor value $x_i \in \mathbb{X}_3=\{11, \ldots, 15\}$. As shown in Figure \ref{fig:2} this construction represents a challenging scenario characterized by different network architectures not generated from our model, and changing across the predictors' values with varying patterns of smoothness and subject-specific variability.

To highlight the possible benefits  provided by our statistical model, we compare performance with a massive univariate approach estimating a flexible logistic regression for each pair of nodes as follows
\begin{eqnarray}
\mathcal{L}(\mathcal{A}_i)_l \mid \pi_l(x_i) \sim \mbox{Bern}\{\pi_l(x_i)\}, \quad \mbox{logit}\{\pi_l(\cdot)\}\sim \mbox{GP}(\bar{\mu},\bar{\sigma}c),
\label{massive}
\end{eqnarray}
for $l=1, \ldots, V(V-1)/2$, where $c$ is the correlation function discussed in Section \ref{method}, $\bar{\mu}$ is the mean function and $\bar{\sigma}$ is a scaling parameter controlling variability. To borrow information across edges, we set $\bar{\mu}$ equal to the log-odds of the empirical edge probabilities computed for each predictor value in $\mathbb{X}$ and let $\bar{\sigma}=10$ to allow flexible deviations in each edge trajectory. Posterior inference under the statistical model in equation \eqref{massive} can be easily performed leveraging the {\rm R} package \texttt{BayesLogit}.

In performing posterior computation under our model we let  $\mu_z=0$, $\sigma^{2}_z=10$, $a=q=2$, $\kappa =0.01$ and set the upper bounds for the latent dimensions at $R=K=5$. We consider $5{,}000$ Gibbs iterations with a burn-in of $1{,}000$ and thin the chains every $4$ samples --- after burn-in. These choices provide good settings for convergence and mixing based on the inspection of the trace-plots for the subject-specific edge probabilities. Posterior computation for our model takes $\sim$16 minutes under a naive {\rm R} (version 3.2.1) implementation in a machine with 8 Intel Core i7 3.4 GHz processor and 16 GB of RAM. Hence, there are substantial margins to reduce computational time. We consider the same MCMC settings when performing posterior inference for the model in \eqref{massive}, obtaining comparable results for convergence and mixing.

\subsection{Inference and predictive performance}
The simulated data set provides a challenging scenario to assess robustness of our methods to model misspecification. We answer this question via posterior predictive checks \citep{gelman2014bayesian} assessing the flexibility of our formulation in characterizing the network summary measures in Figure \ref{fig:2} --- of particular interest in neuroscience \citep{bullmore2009complex}. Calculation of the posterior predictive distributions for these measures is straightforward using the posterior samples of $\pi^{(i)}_l$, $l=1, \ldots, V(V-1)/2$, $i=1, \ldots, n$, and equation \eqref{eq1}.

\begin{figure*}[t]
\centering
\includegraphics[scale=0.57]{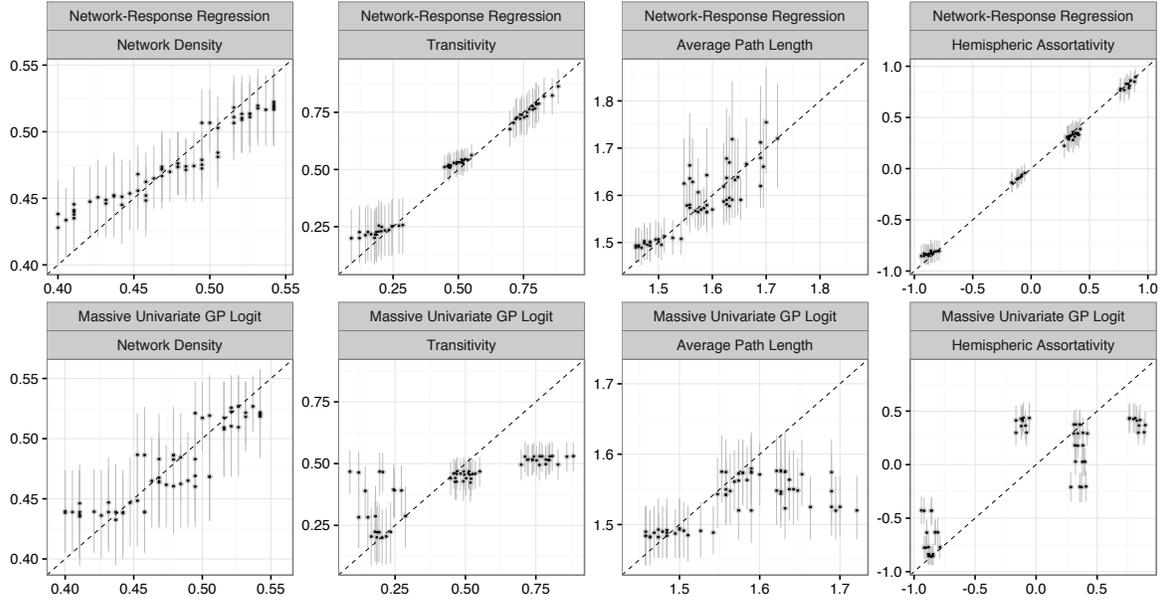}
\caption{Goodness-of-fit assessment for selected network summary measures under the  two different modeling procedures in the simulation study. Upper panels: for the Bayesian network--response regression, plot of the  network summary measures computed from the simulated subjects ($x$-axis) versus their corresponding mean arising from the posterior predictive distribution ($y$-axis). Segments represent the $95\%$ posterior predictive intervals. Lower panels: same quantities from the massive univariate nonparametric logistic regression.}
\label{fig:3}
\end{figure*}

Figure \ref{fig:3} compares the network summary measures computed from the simulated data with their posterior predictive distribution arising from our network--response regression model and the nonparametric massive univariate logistic regressions in \eqref{massive}, respectively. The closer the points in Figure \ref{fig:3} are to the dashed diagonal line, the better the model characterizes the data. According to results in Figure \ref{fig:3}, our formulation achieves general good performance in characterizing the observed networks, even though these data are not generated from a model similar to that described in Section \ref{method}. Differently from our flexible network--response regression, the massive univariate logistic regressions fail to carefully incorporate network structure and subject-specific variability, obtaining worse performance.

Although the substantial dimensionality reduction induced by our low-dimensional factorizations in \eqref{eq2} and \eqref{eq5} provide reassurance against over-fitting issues, we empirically assess this property via out-of-sample edge prediction. In accomplishing this goal, we perform posterior computation under both models holding out, for two networks -- out of four -- in each unique predictor value, those hard-to-predict edges characterized by more evident variability across subjects. For these held-out data -- comprising the $28\%$ of the total number of edges in $A_i$ -- we measure out-of-sample predictive performance via the area under the ROC curve (AUC), computed by predicting the edges with the posterior mean of their corresponding edge probabilities estimated from the training data according to equation \eqref{eq10}. Explicitly incorporating network structure in modeling of each $A_i$, while accounting for subject-specific variability, allows us to obtain also accurate out-of-sample inference with an AUC equal to $0.91$. When performing prediction under the massive univariate logistic regressions in \eqref{massive} we obtain a lower AUC equal to $0.82$. These results confirm the usefulness of our model as a general and flexible procedure to provide accurate inference on global and local changes  in brain networks across traits of interest.

\section{Brain networks and intelligence scores}\label{app}

We apply our model to the dataset \texttt{MRN-114}, which consists of brain structural connectivity data for $n=114$ subjects along with their cognitive ability measured via FSIQ score \citep{jung2007parieto}.  This score ranges from 86 to 144 with 48 unique values observed in our data. In studying the association between intelligence and brain architecture, previous works either focus on detecting the activated brain regions in cognitive control \citep[e.g.][]{leech2012echoes} or study relationships between intelligence and topological network measures \citep[e.g.][]{li2009brain}. We hope to obtain new insights into the neural pathways underlying human intelligence.

In performing posterior computation we consider a total of $5{,}000$ Gibbs samples, setting   $\mu_z= 0$, $\sigma_z^2= 10$, $a=q= 2$, and $\kappa=0.001$. In this case the algorithm required $\sim$3.5 hours. As in the simulation study, trace-plots for the edge probabilities suggest that convergence is reached after 1{,}000 burn-in iterations. We additionally thin the chain by collecting  every 4 samples. Since the latent dimensions $R$ and $K$ are unknown,  we perform posterior computation for increasing $R=K=1,\dots, 6$ and stop when there is no substantial improvement in out-of-sample edge prediction based on the AUC. In computing the AUC, we randomly select 20\% of the edges and hold them out for a randomly chosen 50\% of the subjects.  For these held-out test edges, the  AUC is computed as discussed in the simulation. We repeat the procedure 5 times for each setting and report the average AUC  in Table \ref{tab:AUC-different-R}.  The network-response model having $R=K=5$ provides a good choice for inference and prediction. It is additionally worth noticing how all the AUCs are very high. The reason for this result is that a wide set of brain connections are easier to predict in being either almost always present or absent in the subjects under study.

\begin{table}[t]
\caption{Average AUC computed for the test data at varying $R$ and $K$.\label{tab:AUC-different-R}}
\begin{center}
\begin{tabular}{llllll}
\toprule
$R{=}K{=}1$ & $R{=}K{=}2$ & $R{=}K{=}3$ & $R{=}K{=}4$ & $R{=}K{=}5$ & $R{=}K{=}6$\\
\midrule
0.983 &  0.986 &  0.988 & 0.989 & 0.990 & 0.989\\
\bottomrule
\end{tabular}
\end{center}
\end{table}

In the following subsections we discuss the improved performance of our procedure --- considering  $R=K=5$ --- in relation to our competitor in \eqref{massive}, including in inference, prediction and uncertainty quantification. These relevant improvements are fundamental in refining inference on how network structures change across a continuous trait of interest.

\subsection{Goodness-of-fit and inference on changes in the brain connectivity architecture across FSIQ scores}
\label{inf}

\begin{figure*}[htb]
\centering
\includegraphics[scale=0.57]{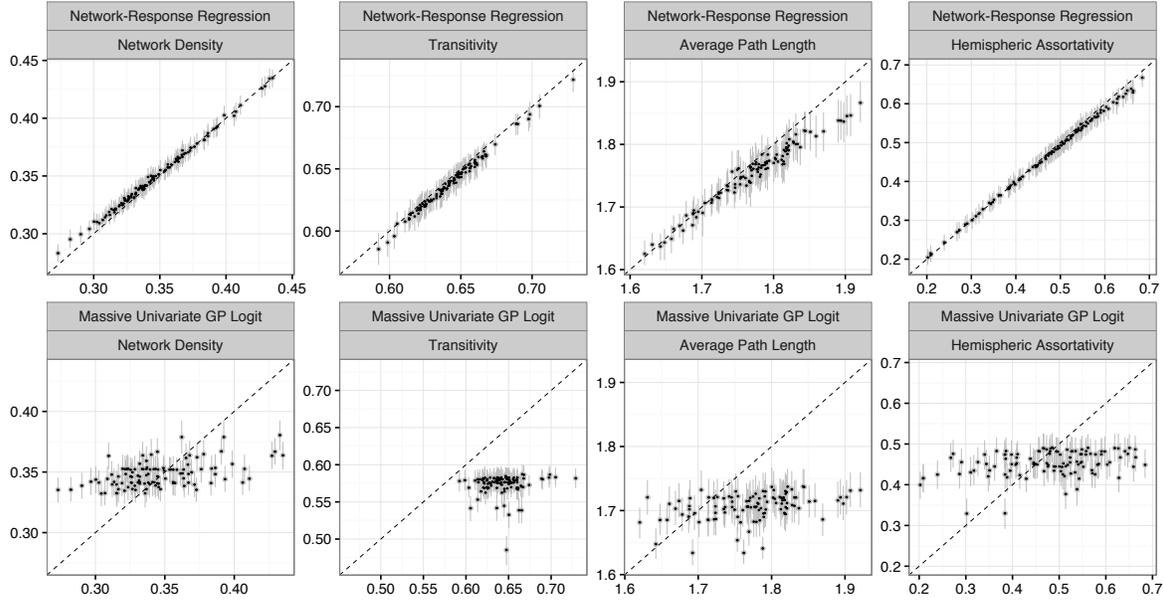}
\caption{Goodness-of-fit assessment for selected network summary measures under the  two different modeling procedures in the application. Upper panels: for the Bayesian network--response regression, plot of the network summary measures computed from the observed subjects ($x$-axis) versus their corresponding mean arising from the posterior predictive distribution ($y$-axis). Segments represent the $95\%$ posterior predictive intervals. Lower panels: same quantities from the massive univariate nonparametric logistic regression.}
\label{fig:4}
\end{figure*}

As discussed in Section \ref{intro} and \ref{simu}, restrictive statistical models for how a brain network architecture changes with a trait can lead to substantially biased inference and conclusions. Hence, prior to presenting our findings, we first assess the performance of our statistical model in characterizing the observed brain network data. Consistent with the analyses in Section  \ref{simu}, this is accomplished via posterior predictive checks for relevant topological properties.  As shown in Figure \ref{fig:4}, our model achieves good performance in characterizing the observed network summary measures, substantially improving over our competitor. This motivates further analyses on how the brain network changes, on average, across FSIQ scores.

\begin{figure}[hbt]
\begin{center}
\includegraphics[width=0.8\columnwidth]{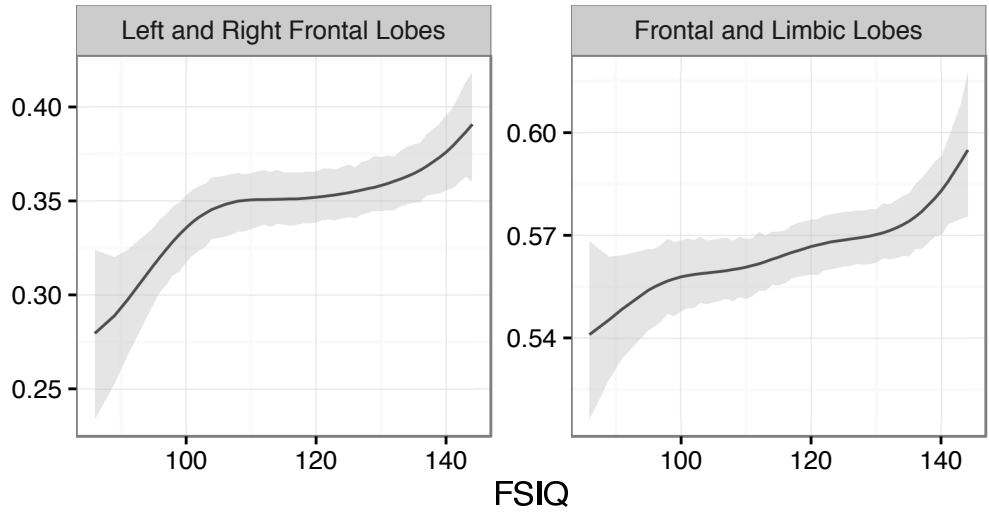}
\caption{{Left plot: average edge probability versus FSIQ for connections linking regions in left and right frontal lobe. Right plot: average edge probability versus FSIQ for connections linking regions in frontal and limbic lobes. Black lines denote the point-wise posterior means and gray areas denote the 95\% highest posterior density intervals from our model.}}
\label{fig:5}
\end{center}
\end{figure}

The cerebrum of the brain is divided into five main anatomical lobes --- named frontal, limbic, occipital, parietal and temporal lobes \citep{kang2012hemispherically}. In order to provide interpretable inference on how the network structure changes with FSIQ, we focus on the posterior distribution for the trajectories of aggregated connectivity patterns considering possible combinations of hemispheric and lobe membership. For example, the left plot in Figure  \ref{fig:5} displays the posterior distribution of the averaged edge probabilities connecting brain regions in  different hemispheres, but belonging both to the frontal lobe.

\begin{figure}[h!]
\begin{center}
\includegraphics[trim=3cm 4.5cm 2.5cm 3.7cm,clip,width=0.4\columnwidth]{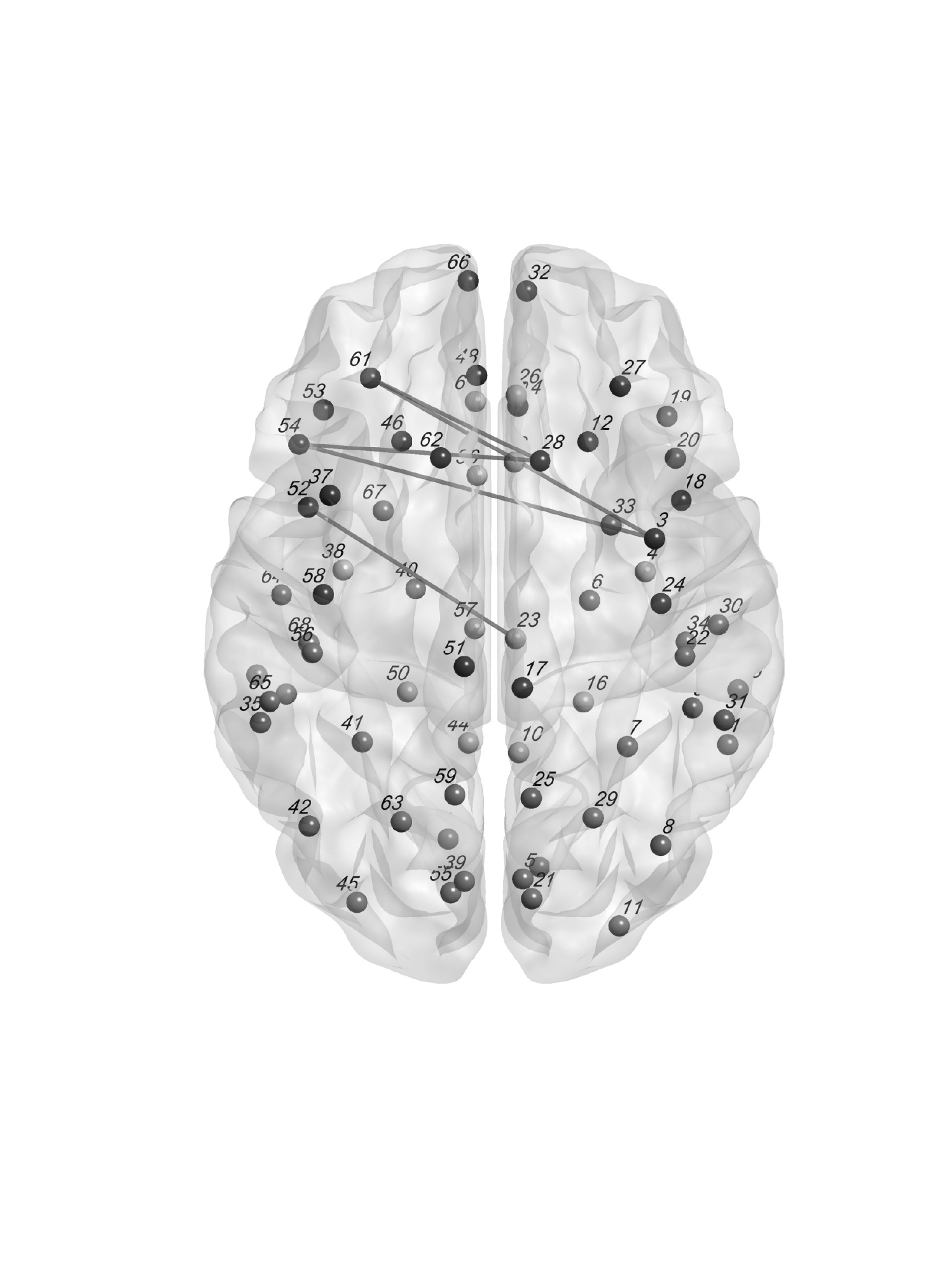}
\caption{{2-D brain network representation showing edges whose trajectories display evident trends across FSIQ -- based on their posterior distributions. Brain regions having the same shade of gray belong to the same anatomical lobe.}}
\label{fig:6}
\end{center}
\end{figure}

Figure \ref{fig:5} shows that the aggregated pathway linking regions in the left and right frontal lobe, as well as the one connecting regions in the frontal lobe with those in the limbic cortex, increase with FSIQ. This result is  in line with  findings on the role of the frontal lobe in intelligence \citep{li2009brain}. To provide insights on local changes in the brain architecture with FSIQ, Figure   \ref{fig:6} highlights the connections whose posterior distributions show evident trends with FSIQ. In particular all the trajectories for the edges highlighted in Figure  \ref{fig:6} significantly increase with FSIQ. Consistent with the results in the left plot of Figure \ref{fig:5}, almost all these edges connect regions in opposite hemispheres but  belonging both to the frontal lobe.

\subsection{Edge prediction and uncertainty quantification}
\label{pred}
Results in Figure \ref{fig:4} are appealing in demonstrating that the substantial dimensionality reduction induced by our model via \eqref{eq2} and \eqref{eq5}, carefully preserves flexibility in characterizing complex brain network structures and their changes with FSIQ. To further investigate the benefits induced by our parsimonious representation we assess performance in edge prediction for a challenging scenario holding out -- for $50\%$ of the subjects -- only the hard-to-predict edges having empirical probability\footnote{The empirical edge probability $\bar{\pi}_{l}$ is defined as $\bar{\pi}_{l}=\sum_{i=1}^{n}\mathcal{L}(A_{i})_l/n$ for each $l=1,\dots,V(V-1)/2$.} $0.2 < \bar{\pi}_{l} < 0.8$. Consistent with the results in the simulation study we obtain an AUC equal to $0.82$, substantially out-performing the AUC of $0.65$ for our competitor.

Another advantage of our flexible Bayesian approach over methods relying on optimization or restrictive hierarchical models is the ability to accurately characterize uncertainty in learning how the brain structure varies with subjects, systematically in relation to a trait and randomly due to unobserved conditions  or measurement errors. We assess performance in uncertainty quantification by evaluating probability calibration in the above prediction task. In particular, we bin the estimated probabilities for the held-out edges in intervals $[0,0.1],(0.1,0.2], \ldots, (0.9,1]$ and -- within each bin -- we calculate the proportion of actual edges among those predicted to have an edge probability within that bin. If the values of these empirical proportions are actually within the bins they refer to, the procedure is well calibrated and properly quantifies uncertainty. According to Table \ref{tab:Probability calibration} our model has good performance in  uncertainty quantification. 

\begin{table*}[h!]
\caption{\footnotesize{Proportion of actual edges among those predicted to have an edge with probability within each bin.\label{tab:Probability calibration}}}
\begin{center}
\tiny
\begin{tabular}{>{\raggedright}m{2.5cm}cccccccccc}
\toprule 
& {[}0, 0.1{]} & (0.1, 0.2{]} &(0.2, 0.3{]} &(0.3, 0.4{]} &(0.4, 0.5{]} &(0.5, 0.6{]} &(0.6, 0.7{]}&(0.7, 0.8{]}&(0.8, 0.9{]} &(0.9, 1{]}\\
\midrule 
Bayesian network--response regression& 0.07 & 0.16 &0.23 &0.34 &0.42 & 0.53 & 0.62& 0.70 & 0.80&0.91\\
Massive univariate GP logistic regression&0.23 & 0.27 &0.33 &0.39 &0.45 &0.50 &0.59& 0.62&0.63 &0.70\\
\bottomrule
\end{tabular}
\end{center}
\end{table*}

\section{Conclusion}\label{conclusion}
Motivated by a neuroscience study, we developed a novel model to flexibly infer changes in a network-valued random variable with a continuous trait.  
The simulation study and the application to learn variations in the brain connectivity architecture across FSIQ show substantial improvements in inference, edge prediction and uncertainty quantification.

Although we focus on a single trait, the method is trivially generalized to accommodate multiple traits of an individual. Moreover our formulation can be easily adapted to incorporate directed networks via two subsets of latent coordinates --- for each brain region --- modeling outgoing and incoming edges, respectively. Our procedure also has a broad range of possible applications in social science and econometrics. 

Although our initial results suggest that binary brain networks already contain valuable information about the individual's brain structure, future research generalizing the proposed model to account for weighted edges, may benefit from the additional information contained in the fiber counts. 

\section*{Acknowledgements}

We thank Joshua T. Vogelstein for getting us interested in statistical and computational methods for analysis of brain networks. We are also grateful to the Associate Editor and the referees for their valuable comments on a first version of this manuscript.

\section*{Funding}

This work was partially supported by the grant N00014-14-1-0245 of the United States Office of Naval Research (ONR), and grant CPDA154381/15 of the University of Padova, Italy.

\newpage
\begin{center}
{\large\bf SUPPLEMENTARY MATERIAL}
\end{center}

\section*{Posterior computation}
Given the priors defined in equations (3)--(8) and based on the P\'olya-Gamma data augmentation for Bayesian logistic regression, the Gibbs sampler for our network-response regression model in (1)--(2) alternates between the following steps. 
\begin{itemize}
\item Update the P\'olya-Gamma augmented data for each pair of brain regions $l$ in every subject $i$, from 
\[
\omega_{l}^{(i)}\mid - \sim\mbox{PG}\left(1,Z_l+\sum_{r=1}^R Y^{(i)}_{vr}Y^{(i)}_{ur}\right),
\]
for every $l=1,\dots,V(V-1)/2$ and $i=1,\dots,n$.
\item Sample --- for each subject $i=1, \ldots, n$ --- the latent coordinates for her nodes comprising the $V\times R$ matrix $Y^{(i)}$. We accomplish this by block updating the elements in each row $Y_{v\cdot}^{(i)}$, $v=1, \ldots, V$ of $Y^{(i)}$ --- representing the $R$ coordinates of node $v$ in subject $i$ --- given all the others $u \neq v$. Recalling equations (1)--(2) we can obtain the full conditional posterior distribution for $Y_{v\cdot}^{(i)\intercal}$, by recasting the problem as a Bayesian logistic regression with  $Y_{v\cdot}^{(i)\intercal}$ acting as a coefficient vector. In particular, let 
\begin{align*}
\mathcal{L}(\mathcal{A}_{i})_{(v)}\mid\pi_{(v)}^{(i)} & \sim\mbox{Bern}\{\pi_{(v)}^{(i)}\},\\
\mbox{logit}\{\pi_{(v)}^{(i)}\} & =Z_{(v)}+Y_{(-v)}^{(i)}Y_{v\cdot}^{(i)\intercal},
\tag{11}
\end{align*}
where $Y_{(-v)}^{(i)}$ denotes the $(V-1)\times R$ matrix obtained by removing the $v$th row of $Y^{(i)}$, while $\mathcal{L}(\mathcal{A}_i)_{(v)}$ and $Z_{(v)}$ are $(V-1)\times 1$ vectors obtained by stacking elements $\mathcal{L}(A_i)_l$ and $Z_l$ for all the $l$ corresponding to pairs $(u,w)$ such that $u=v$ or $w=v$, with $u>w$ and ordered consistently with equation (11). Recalling equations (4)--(5), the prior for $Y_{v\cdot}^{(i)\intercal}$ is $\mbox{N}\{W(x_{i})^{\intercal}G_{v\cdot}^{\intercal},\mbox{I}_{R}\}$, with $G$ the $V \times K$ matrix of coefficients, $W(x_{i})$ the $K \times R$ matrix containing the values of the basis functions at $x_i$ and $\mbox{I}_{R}$ the $R \times R $ identity matrix.  Hence, the P\'olya-Gamma data augmentation for the model (11) ensures that the full conditional for each row of $Y^{(i)}$ is
\[
\begin{array}{l}
Y_{v\cdot}^{(i)\intercal}\mid - \sim \mbox{N}_{R}\{\mu_{v}^{(i)},\Sigma_{v}^{(i)}\}, \quad v=1,\dots,V,
\end{array}
\]
with
\[
\begin{array}{l}
\Sigma_{v}^{(i)}=\{\mbox{I}_{R}+Y_{(-v)}^{(i)\,\intercal}\Omega_{(v)}^{(i)}Y_{(-v)}^{(i)}\}^{-1},\\
\mu_{v}^{(i)}=\Sigma_{v}^{(i)}\{ W(x_{i})^{\intercal}G_{v\cdot}^{\intercal}+Y_{(-v)}^{(i)\,\intercal}\psi^{(i)}_{v}\},\\
\psi^{(i)}_{v}=\mathcal{L}({A}_i)_{(v)}-0.5\cdot\mathbf{1}_{V-1}-\Omega_{(v)}^{(i)}Z_{(v)},
\end{array}
\]
with $\Omega_{(v)}^{(i)}$ the $(V-1)\times(V-1)$ diagonal matrix with entries obtained by stacking the P\'olya-Gamma augmented data consistently with (11). 
\item Sample each shared similarity score $Z_{l}$, $l=1, \ldots, V(V-1)/2$ from its Gaussian full conditional
\[
Z_{l} \mid - \sim \mbox{N}\left(\mu_{l}^{Z},\sigma_{l}^{Z}\right),
\]
where $\sigma_{l}^{Z}=1/(\sigma^{-2}_z+\sum_{i=1}^{n}\omega_{l}^{(i)})$ and $\mu_{l}^Z=\sigma_{l}^{Z}[\sigma^{-2}_z\mu_z+\sum_{i=1}^n\{\mathcal{L}({A}_i)_{l}-1/2-\omega_{l}^{(i)} Y_{v\cdot}^{(i)}Y_{u\cdot}^{(i)\intercal} \}]$, with $v$ and $u$ the nodes corresponding to pair $l$.
\item Update each basis function $W_{kr}(\cdot)$, $k=1, \ldots, K$ and $r=1, \ldots, R$ from its full conditional posterior. In particular, our Gaussian process prior (6) for the basis functions implies that 
\begin{eqnarray*}
\begin{pmatrix}W_{kr}(x_{1}^{\star})\\
\vdots\\
W_{kr}(x_{n^*}^{\star})
\end{pmatrix} & \sim & \mbox{N}_{n^*}\left\{\begin{array}{ll}
\begin{pmatrix}0\\
\vdots\\
0
\end{pmatrix}, & C\end{array}\right\},
\end{eqnarray*}
where $(x_{1}^{\star},\dots,x_{n^*}^{\star})$ are the unique values of $(x_{1},\dots,x_{n})$ and $C$ is the Gaussian process covariance matrix with $C_{ij}=c(x_{i}^{\star},x_{j}^{\star})$. Hence, in updating $\{W_{kr}(x_{1}^{\star}), \ldots, W_{kr}(x^{\star}_{n^*}) \}^{\intercal}$, let $D=\mbox{diag}\{\sum_{i}{\bf 1}(x_i=x_{1}^{\star}), \ldots, \sum_{i}{\bf 1}(x_i=x^{\star}_{n^*})\}$, and
\[
\begin{array}{ccccccc}
\hat{Y}_{\cdot r} & = & \begin{pmatrix}\sum_{i:x_{i}=x_{1}^{\star}}Y_{1r}^{(i)}\\
\vdots\\
\sum_{i:x_{i}=x_{n^{*}}^{\star}}Y_{1r}^{(i)}\\
\sum_{i:x_{i}=x_{1}^{\star}}Y_{2r}^{(i)}\\
\vdots\\
\sum_{i:x_{i}=x_{n^{*}}^{\star}}Y_{2r}^{(i)}\\
\vdots\\
\sum_{i:x_{i}=x_{1}^{\star}}Y_{Vr}^{(i)}\\
\vdots\\
\sum_{i:x_{i}=x_{n^{*}}^{\star}}Y_{Vr}^{(i)}
\end{pmatrix} & , & \hat{W}_{\cdot r} & = & \begin{pmatrix}W_{1r}(x_{1}^{\star})\\
\vdots\\
W_{1r}(x_{n^{*}}^{\star})\\
W_{2r}(x_{1}^{\star})\\
\vdots\\
W_{2r}(x_{n^{*}}^{\star})\\
\vdots\\
W_{Kr}(x_{1}^{\star})\\
\vdots\\
W_{Kr}(x_{n^{*}}^{\star})
\end{pmatrix}.\end{array}
\]
Standard conjugate posterior analysis provides the following full conditional
\[
\begin{array}{l}
\hat{W}_{\cdot r} \mid - \sim \mbox{N}_{Kn^*}\left(\mu_{r}^{W},\Sigma_{r}^{W}\right),\end{array}
\]
for each $r=1,\dots,R$, with
\[
\begin{array}{l}
\Sigma_{r}^{W}=\left(\mbox{I}_{K}\otimes C^{-1}+G^{\intercal}G\otimes D\right)^{-1},\\
\mu_{r}^{W}=\Sigma_{r}^{W}(G^{\intercal}\otimes \mbox{I}_{n^*})\hat{Y}_{\cdot r}.
\end{array}
\]
\item Conditioned on the hyperparameters $\tau_{k}$, the Gaussian prior on the elements of $G$ in equation (7) yields the following full conditional for each row of $G$:
\[
G_{v\cdot}^{\intercal} \mid - \sim \mbox{N}_K(\mu_{v}^{G},\Sigma_{v}^{G}),
\]
for each $v=1,\dots,V$, with
\[
\begin{array}{l}
\Sigma_{v}^{G}=\left\{\tau+\sum_{i=1}^{n}W(x_{i})W(x_{i})^{\intercal}\right\}^{-1}\\
\mu_{v}^{G}=\Sigma_{v}^{G}\left\{\sum_{i=1}^{n}W(x_{i})Y_{v\cdot}^{(i)\intercal}\right\}
\end{array}
\]
where $\tau=\mbox{diag}(\tau_1,\tau_2,\ldots,\tau_K)^{\intercal}$. 
\item The global shrinkage hyperparameters are updated as 
\[
\tau_{k} \mid -  \sim\mbox{Ga}\left(aq^{3(k-1)}+\frac{V}{2},q^{2(k-1)}+\frac{1}{2}\sum_{v=1}^{V}G_{vk}^{2}\right),
\]
for each $k=1, \ldots, K$.
\item Update the subject-specific edge probabilities by applying equation 
\[
\pi_l^{(i)}=\left\{1+\exp(-Z_l-\sum_{r=1}^R Y^{(i)}_{vr}Y^{(i)}_{ur})\right\}^{-1},
\]
to the posterior samples of $Z_l$ and $Y^{(i)}$ for each $l=1, \ldots, V(V-1)/2$ and $i=1, \ldots, n$.
\end{itemize}  

\newpage
\bibliographystyle{natbib}

\end{document}